# A Localization Strategy Based on *N*-times Trilateral Centroid with Weight

Tie Qiu, Yu Zhou, Feng Xia*, Naigao Jin, Lin Feng
School of Software, Dalian University of Technology, Dalian 116620, China
Email: qiutie@dlut.edu.cn; zhouyujoe@126.com; f.xia@ieee.org
* Corresponding author; Email: f.xia@ieee.org

*Abstract*—**Localization based on received signal strength indication (RSSI) is a low cost and low complexity technology, and it is widely applied in distance-based localization of wireless sensor networks (WSNs). Error of existed localization technologies is significant. This paper presents the *N*-times trilateral centroid weighted localization algorithm (NTCWLA), which can reduce the error considerably. Considering the instability of RSSI, we use the weighted average of many RSSIs as current RSSI. To improve the accuracy we select a number of (no less than three) reliable beacon nodes to increase the localization times. Then we calculate the distances between reliable beacon nodes and the mobile node using an empirical formula. The mobile node is located *N* times using the trilateral centroid algorithm. Finally, we take the weighted average of the filtered reference coordinates as the mobile node's coordinates. We conduct experiments with the STM32W108 chip which supports IEEE 802.15.4. The results show that the proposed algorithm performs better than the trilateral centroid algorithm.**

## I. INTRODUCTION

Wireless sensor networks (WSNs) are rapid self-organized and distributed networks, which are composed of many battery-powered, low-cost wireless sensor nodes deployed in monitoring area [1-3]. Node-localization technology is one of the supporting technologies of WSNs and many location-based works won't run without location information of sensor nodes. Some classic localization algorithms may not be feasible for WSNs due to the limitations of ability and battery energy of nodes. Therefore, we focus on devising a low-power and high-precision localization algorithm. Localization methods based on received signal strength indication (RSSI) can satisfy those requirements and do not need additional hardware.

According to whether it requires to measure the distances between nodes in WSNs, localization algorithms are divided into distance-dependent and distance-independent localization algorithms. Distance-dependent localization algorithms need to measure the actual distances or orientation between adjacent nodes, and then use the measured data to locate unknown nodes. Distance-independent localization algorithms do not need measure the actual distances or orientation between nodes, but use the estimated distances between nodes to calculate unknown nodes' positions. The nodes whose locations are known are beacon nodes. The RSSI localization algorithm [4] calculates the distances between nodes by measuring the signal attenuation. However, the stability of RSSI is poor so the distances aren't accurate, which results in the inaccuracy of localization. Time of arrival (TOA) [5] measures the distances between nodes by calculating the transmission time of one kind of signal. Time difference of arrival (TDOA) [6] measures the distances by calculating the arrival-time interval of two different signals. Angle of Arrival (AOA) [7] measures the arrival-direction of signals with antenna array, and then calculates the relative orientation or angle according to the measured arrival-direction. Finally, it uses the triangulation method to calculate the mobile node's position. These three methods require additional hardware support, which increases the spending. Methods in [8-10] are all distance-independent localization algorithms. Centroid algorithm [8] uses the geometric center of a mass of beacon nodes as the unknown nodes' locations. This method is simple but the error is large, so it is impractical. DV-Hop algorithm [9] uses hop counts to denote distance and the error is great. The unknown node calculates the distances with beacon nodes by measuring the hop counts with beacon nodes, and then uses the measured distance to locate itself. The approximate point-in-triangulation test algorithm (APIT) [10] locates the unknown node via continuously identifying whether the unknown node is within the triangle area, which is structured by three beacon nodes. This method can only calculate the position roughly. Error of the mentioned localization methods is big so they are not suitable for accurate localization. Methods in [11-13] are all put forward based on trilateration. Localization accuracy can be improved if we take weight into account for trilateral localization algorithm [11] or use a modified model to correct the measured RSSI [12-13]. It is a new method to combine DV-Distance with centroid localization algorithm and at the same time take weight into account [14]. However, the accuracy is constrained by the accuracy of DV-Distance. Using many groups of historical RSSI to calculate current RSSI can reduce the influence of instability of RSSI [15], but it doesn't consider the weight of historical value. We improve it by taking weight of historical value into account to calculate current RSSI. The authors of [16] use the sum of the measured distances' reciprocal instead of the reciprocals of the measured distances' sum as the weight and puts forward correction-factor to avoid the information inundate phenomenon. This new method of calculate weight ensures that with the increase of the distances between nodes the weight is decreasing. Received signal strength difference algorithm (RDWCL) [17] defines different weights for different beacon nodes based on RSSI. Then it uses the weighted centroid algorithm to achieve location. Beacon nodes' layout has important influence on the localization accuracy and the best layout of beacon nodes is equilateral triangle. The authors of [18] improve the centroid algorithm to ellipse centroid localization algorithm and at the same time combine it with error-factor and precision-factor. The reliability of the beacon node is high if it is close to mobile node and the accuracy of localization is high if localization is performed with reliable beacon node. Triangle or polygon centroid localization algorithm selects the three or more beacon nodes which are closest to the mobile node to locate the mobile node [19]. The localization accuracy of taking the LQI (link quality indicator) value into account during the weighted centroid localization is better than the weighted centroid localization algorithm [20]. Method mentioned in [21] divides the block into different blocks according to the relative positions of beacon nodes and mobile node, and then defines different priorities for different blocks. Different priorities represent different weights. This is a new way to get weight for beacon nodes. However, the accuracy of this method is significantly influenced by the measured accuracy of RSSI. Polygon algorithm [22] is better than trilateration. The way to get better parameters for empirical formula which converts RSSI to distance is detailed described in [22] and we propose a new method to get parameters which is better than it. The authors of [23] deal with the wireless sensor networks based on IEEE802.15.4 protocol and analyze three indoor environment localization methods based on RSSI: trilateration, minimax algorithm, and maximum likelihood estimation. The experiment results show that in the indoor environment the trilateration shows excellent properties and the availability is good.

There are many efficient algorithms but many of them require additional hardware support. Localization methods based on RSSI do not need any additional hardware support and do not have to keep synchronous of the network. Compared with distance-independent algorithms distance-dependent algorithms are more accurate. Trilateral centroid algorithm uses RSSI to calculate the distances between nodes and it is a distance-dependent algorithm. And it is better than trilateration and centroid positioning method. In order to improve the localization accuracy, we propose the *N*-times trilateral centroid weighted localization algorithm (NTCWLA, *N* denotes the times of using trilateral centroid algorithm) [24]. We select *n* (n≥3) reliable beacon nodes and then combine any three of them to calculate the reference coordinates of mobile node with trilateral centroid algorithm. At the same time we add weight to each reference coordinates. The method of calculating the weight is basically the same with [16]. Next we use the weighted average of reference coordinates to filter out the reference coordinates. Finally, we take the weighted average of the filtered reference coordinates as the mobile node's coordinates. In order to improve the reliability of localization and reduce the influence of the instability of RSSI, mobile node selects reliable beacon nodes periodically and takes the weighted average of multiple RSSIs as the node's current RSSI. Getting more accurate RSSI is a prerequisite for localization and localization needs a better empirical formula. Localization accuracy is increasing with the increase of the reliability of beacon nodes. Value of *n* affects the reliability of beacon nodes.

The rest of this paper is organized as follows: In Section II, we discuss some problems of localization. Section III describes the method of NTCWLA. Section III.A presents the way of getting empirical formula. Section III.B is about the selection of reliable beacon nodes. How to adjust the localization period is discussed in Section III.C. Section III.D gives the main idea of NTCWLA. The implementation of the new algorithm is described in Section IV. Section V presents all experiments and the experimental results. Finally, we conclude the paper in Section VI.

## II. PROBLEM STATEMENT

Indoor transmission loss of electromagnetic wave not only relates with transmission distance, but also relates with the obstacles in the path. The transmission loss model is detailed described in [8].

$$P(d)[dBm] = P(d_0)[dBm] - 10\eta \log(\frac{d}{d_0}) - \zeta \quad (1)$$

$d_0$ is the reference distance. $P(d_0)$ is the RSSI when transmission distance is $d_0$. $d$ is the actual distance. $P(d)$ is the RSSI when distance is $d$. $\zeta$ is the environment factor which has nothing to do with the transmission distance, but relates to the obstacle in the environment, movement of people, temperature and humidity. $\eta$ is the index of path loss which relates to the structure and material of building. The function relationship between $d$ and $P(d)$ has a close relationship with $\eta$, $\zeta$ and the reference point ($d_0$, $P(d_0)$). These parameters determine the accuracy of the calculated distances between nodes, which affect the localization accuracy, so getting better parameters can be as the first important step for localization.

After getting the parameters of Formula (1), obtaining more reliable RSSIs of beacon nodes is the next important step for localization. In actual the fluctuation of RSSI causes larger error of localization. Generally, we take the weighted average of multiple groups of historical RSSI as current RSSI to reduce the error, and then select several beacon nodes to locate the mobile node. Here we meet two problems.

We should take how many packets. Firstly, it takes a long time to receive enough packets from beacon nodes if we take too many packets as current RSSI. However, localization requires real time. Localization loss the attribute of real time if we take too many packets as current RSSI and it reduces the accurate of localization. Secondly, RSSI is volatile and the accurate of current RSSI is reduced if we just take several packets as current RSSI. So it influences the accurate of the measured distance between nodes and increases the localization error. Last, the packet number affects the reliable beacon nodes' number of a localization period. A beacon node can be a reliable beacon node only if the mobile node receives enough packets from it in a localization period, so more packets lead to less reliable beacon nodes. If the localization period is fixed, we should set a better value for the group number of RSSI to improve the localization accuracy.

Figure 1 depicts the relationship between actual distances and measured RSSIs. The measured RSSIs are the average values of RSSI of many packets in a fixed location. The number of packets $TN$ is 5, 10, 15, 20 and 25. Figure 1 shows that with the increase of $TN$ the average value of RSSI is increasing. The trend of the relationship curve between distances and RSSIs is negative exponential approximately. Figure 2 depicts the relationship between actual distances and measured distances. The measured distances are calculated by putting the average values of RSSI in Figure 1 into an empirical formula. The measured distances of $TN=20$ and $TN=25$ are more accurate than $TN=5$ and $TN=10$, which can explain that the average value of RSSI of $TN=20$ and $TN=25$ are more accurate in Figure 1. In actual the rate of communication between nodes is limited and mobile node demands to communicate with many beacon nodes, so it will take a long time if mobile node receives 20 or 25 packets from each reliable beacon node. The error is large if we locate the mobile node after a long time. However, the localization accuracy is very poor if we only collect 5 packets from a beacon node.

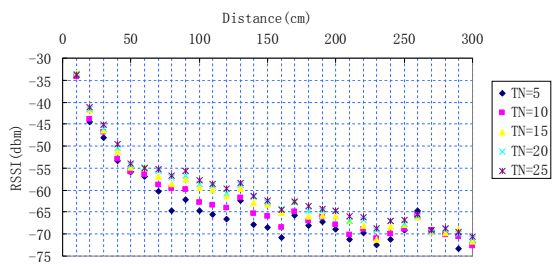

Figure 1. Actual distances and average values of RSSI

The number of reliable beacon nodes isn't the more the better, the reliability and accuracy of selected reliable beacon nodes decide the localization accuracy. First it increases the times of using the trilateral centroid algorithm and increases the computation with more reliable beacon nodes, which doesn't meet the requirement of energy saving. Second the localization accuracy will be decreased if locate with some reliable beacon nodes, whose reliability is low. However, it will be same with the trilateral centroid algorithm if localization with less reliable beacon nodes and cannot improve the localization accuracy. In actual use the number of reliable beacon nodes is decided by the layout of beacon nodes.

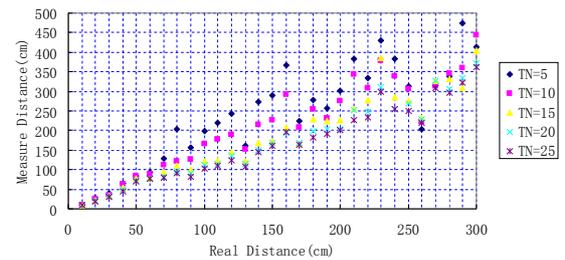

Figure 2. Actual distances and measured distances

Figure 3 illustrates that the reliable beacon nodes' number affects the localization accuracy. In Figure 3, the measured position 1 indicates the measured position with 3 reliable beacon nodes, the measured position 2 denotes the measured position with 5 reliable beacon nodes, and the measured position 3 indicates the measured position with 10 reliable beacon nodes. The amount of computation with 3 reliable beacon nodes to locate is less than that with 5 reliable beacon nodes to locate, but the localization error is great. The amount of computation is more when localization with 10 reliable beacon nodes. But the localization error is increased because some beacon nodes' reliability is decreasing. The reliability and the number of reliable beacon nodes affect the localization accuracy, so we should get a balance between reliability and number of reliable beacon nodes, namely gets a better value for $n$.

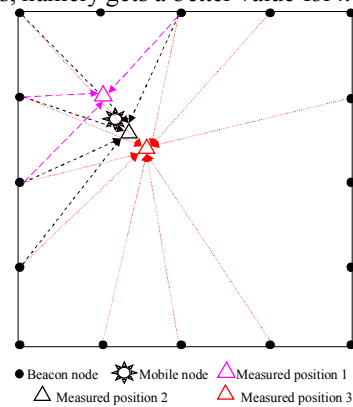

Figure 3. Localization accuracy and number of reliable beacon nodes

## III. N-TIMES TRILATERAL CENTROID WEIGHTED LOCALIZATION

### A. Empirical Formula

Empirical formula is a basic condition to localization, Formula (1) is a theoretical model. We need to get the parameters based on the actual environment. To achieve the empirical formula we experiment to collect a large number of data under the actual environment. In the experiment the mobile node's location is fixed and all beacon nodes are removable. The send-power of all beacon nodes is same and all beacon nodes communicate with the mobile node in same distance. The mobile node saves the RSSIs and the corresponding distances of beacon nodes. In each fixed distance the communication lasts a certain time (e.g. 1*min*). End of the experiment we deal with the saved data to gain empirical formula.

After the experiment, firstly, we filter out the signal whose RSSI is too small, such as less than -70*dbm*. Secondly, we calculate the average value of RSSI of a fixed distance and save the average value with the corresponding distance. Formula (1) shows that $d$ has a *log* relationship with $P(d)$, suppose $x = \log^d$, so Formula (1) is converted to (2)

$$P(d) = p_1 \cdot x + p_2 \quad (2)$$

$x$ is independent variable and $P(d)$ is dependent variable. We deal with the saved data to ensure the value of $p_1$ and $p_2$, and then use $x = \log^d$ to replace $x$ and inverse the relationship between $d$ and $P(d)$. Last we get Formula (3).

$$d = e^{(\frac{P(d) - p_2}{p_1})} \quad (3)$$

By putting the average values of RSSI calculated previously into Formula (3) we can get the measured distances. In the actual environment the measured RSSI is less than the theoretical value when $d$ is small and the measured RSSI is more than the theoretical value when $d$ is large. So when $d$ is small the measured distance is less than the actual distance and in one position the negative deviation is maximal. When $d$ is large the measured distance is bigger than the actual distance and in one position the positive deviation is maximal. The beacon node is more reliable if it is closer to mobile node. We should minimize the negative deviation to improve the localization accuracy while can't overly enlarge the positive deviation. We adjust the value of *smooth* during data processing to get better parameters. The default value of *smooth* is 1. With the increasing of smooth the negative deviation is decreased and the positive deviation is increased. So we should get a balance between positive deviation and negative deviation.

It can be verified that Formula (3) gets the best parameters when $p_1$ gets the minimal value and at the same time $p_2$ gets the maximal value.

If we suppose the group number of $P(d)$ and $d$ is *num*, then the possible value of *smooth* is $i$ ( $0 < i \leq num, i\%2 = 1$ ). By adjusting *smooth* we can get many sets of $p_1$ and $p_2$, respectively, expressed by $p_{1i}$ and $p_{2i}$. The criteria of selecting the optimal values of $p_1$ and $p_2$ is: if $p_{1j}$ gets the minimal value and $p_{2j}$ gets the maximal value, the optimal values of $p_1$ and $p_2$ are $p_1 = p_{1j}$, $p_2 = p_{2j}$; if $p_{1j}$ gets the minimal value, $p_{2j}$ isn't the maximal value, but $p_{2m}$ is the maximal value, the optimal values of $p_1$ and $p_2$ are

$$p_1 = \frac{p_{1j} + p_{1m}}{2} \quad (4)$$

$$p_2 = \frac{p_{2j} + p_{2m}}{2} \quad (5)$$

### B. Excellent Beacon Nodes

The selection of reliable beacon nodes has an important influence on the accuracy of localization. The measured RSSIs are unstable if reliable beacon nodes and mobile node are far apart, which leads to the inaccuracy of the measured distances. Thereby, it will affect the localization accuracy. Reliable beacon nodes must ensure that the deviation between the actual distances and the measured distances is small. The measured distances are calculated by taking the current RSSIs into Formula (3) and the accuracy of current RSSIs determines the deviation. In order to ensure the accuracy of current RSSIs, reliable beacon nodes must ensure that the mobile node can receive a certain number of packets from them and the RSSI of these packets must be bigger than a threshold. For example, the packets' number is no less than 5 and the RSSI of all these packets must be bigger than -55*dbm*. Beacon node can't be a reliable beacon node even if the mobile node receives some packets from it, but the number of the packets is less then 5. If there are many beacon nodes meeting above conditions we should do further optimized selection based on RSSI to minimize the computation and improve the reliability of beacon nodes. For instance, we just select 5 or 6 reliable beacon nodes from 10 or 20 beacon nodes to locate the mobile node. If the number of beacon nodes which meet above conditions is less, for example, less than 5, we do not need to select better beacon nodes and just use them to locate the mobile node. If the frequency of this case is very high, it indicates the localization period is too short, so we need to increase the localization period to increase reliable beacon nodes.

## C. Localization Period Adjustment

Localization period affects the number and quality of reliable beacon nodes, thus it can affect the localization accuracy. Localization will lose the attribute of real time if its period is too long. To a moving object if localization loses the attribute of real time it will lose accuracy. However, if the localization period is very short, the group number of RSSIs will be very small and reliable beacon node will be fewer in a localization period, so the localization accuracy will be very bad. This can be seen from Figure 2, where the localization error of $TN$=5 is bigger than the localization error of $TN$=20. $TN$=5 means short localization period and $TN$=20 indicates a logical localization period. A logical localization period can ensure that there are enough reliable beacon nodes in a localization period. We adjust the localization period to a better value according to the number of reliable beacon nodes in the process of localization during a certain long time. In the beginning, we initialize the localization period with a small value, then periodically check the reliable beacon nodes' number of a localization period and adjust the localization period based on the check result. Figure 4 is a simple flow chart, which describes the adjustment of localization period.

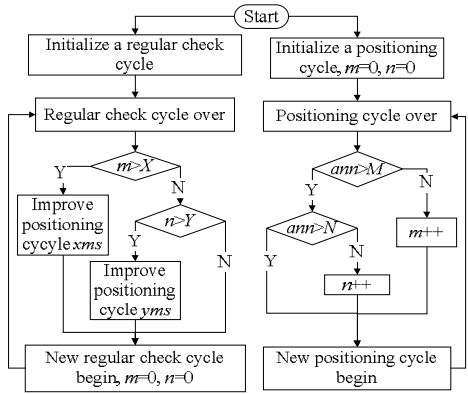

Figure 4. Adjustment of localization period

Note: The description of the symbols in Figure 4. $M$, $N$, $X$, $Y$, $x$ and $y$ are constants. $M$ and $N$ are used to check whether the number of beacon nodes meets a certain requirement, in which $M<N$. $m$, $n$ and $ann$ are variables, $ann$ records the number of reliable beacon nodes in a localization period. $m$ and $n$ are used to record the times of the number of reliable beacon nodes less than $M$ and $N$ during a regular check period, respectively. $X$ and $Y$ are used to check whether the value of $m$ and $n$ meets a certain requirement. System will adjust the localization period dynamically based on the checkup. $x$ and $y$ is the dynamic adjustment range of localization period.

## D. Localization Method

Trilateration is an ideal model. The three circles intersect at a point when there is no error of the measured distances between nodes and this point is the location of the mobile node. However, in actual situation the error of the measured RSSI is existent. The three circles won't intersect at a point because of the measured error, but usually they intersect at a region. Taking the centroid of the region as the location of mobile node is the idea of trilateral centroid algorithm. NTCWLA is an expansion of trilateral centroid algorithm. First we select $n$ ($n \geq 3$) reliable beacon nodes from all beacon nodes, then we calculate the distances between reliable beacon nodes and mobile node with Formula (3). Next we combine any three of the $n$ reliable beacon nodes to calculate the location of mobile node with the trilateral centroid algorithm and this algorithm is executed $N$ ( $N = C_n^3$ ) times. After that we get $N_1$ ( $N_1 \leq N$) reference coordinates of mobile node and use the weighted average of the $N_1$ reference coordinates to filter out the reference coordinates which have large deviation with it. For example, the value of deviation is more than 20$cm$. Finally we take the weighted average of the filtered reference coordinates as the mobile node's coordinates.

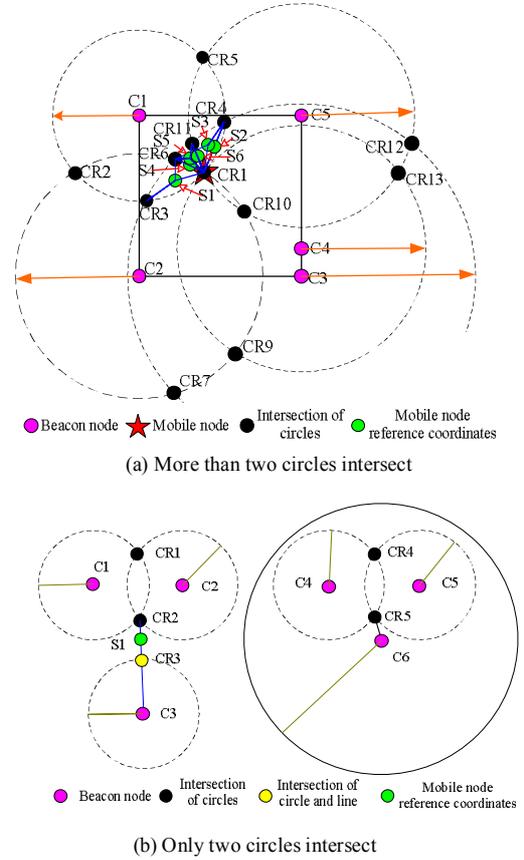

Figure 5. Relationship of circles of NTCWLA

Figure 5 (a) and Figure 5 (b) describe the relationship of the circles, whose centers are the locations of beacon nodes and radius are the measured distances between mobile node and beacon nodes. There are five beacon nodes in Figure 5 (a) so there are five circles C1, C2, C3, C4 and C5. Figure 5 (b) is the complementarities of Figure 5 (a). The possible relationship between any three circles is analyzed as follows: The three circles will intersect at a common point if the measured distances between mobile node and reliable beacon nodes are without any error. The common point is the actual location of the mobile node. However, the error of the measured RSSI is existent and the empirical formula also has error. The error cause the measured distance less than or more than the actual distance, which increases the relationship between the three circles.

**Case 1:** The three circles have a common point.

In Figure 5 (a) C1, C2 and C4 intersect at one point CR1, which is a reference coordinates of mobile node. In Figure 5 (a) C1, C2, C5 and C1, C4, C5 and C2, C4, C5 are all consistent with case 1 and the common point of them is CR1. The common point of the three circles is the reference coordinates of mobile node. Case 1 is an ideal situation and in actual it is rare.

**Case 2:** The three circles determine a region.

In Figure 5 (a) C1, C2 and C3 determine a region, which is consistent with case 2. The region determined by points CR1, CR3 and CR6 and the centroid of this region is S1, which is a reference coordinates of the mobile node. In Figure 5 (a) S3 is the centroid of the region which is determined by C1, C3 and C5; S5 is the centroid of the region which is determined by C2, C3 and C5. The way of determining the reference coordinates of mobile node for case 2 is as follows: we take C1, C2 and C3 as an example. The intersections of C1 and C2 are CR1 and CR2 and we take the point CR1 which is in the test area. The intersections of C1 and C3 are both in the test area and we select the point CR3 which is close to the center of C2. We take the point CR6, which is in the test area, the one intersection of C2 and C3. Last we take the centroid S1 of the area which is determined by CR1, CR3 and CR6 as the mobile node's reference coordinates. If the relationship of the three circles meets case 2, we first check whether the two intersections of two circles are in the test area, if there is only one in, then take it; if the two intersections are both in the test area, we take the one which is close to the center of the third circle.

**Case 3:** The three circles determine a line.

In Figure 5 (a) the relationship of C1, C3 and C4 meets case 3, the reference coordinates of the mobile node are the mid-point S2 of CR1 and CR4. In Figure 5 (a) the reference coordinates S4 is determined by C2, C3 and C4, the node S6 is determined by C3, C4 and C5. The way to determine the reference coordinates of mobile node of case 3 is as follows: we select C1, C3 and C4 as an example. CR3 and CR4 are the intersections of C1 with C3. These two points are both in the test area and we take the point CR4 which is closer to the center of C4. Then we take the intersection CR1 of C1 and C4, last we take the mid-point S2 of CR1 and CR4 as the reference coordinates of mobile node.

**Case 4:** Only two circles have intersection.

There are two cases meet case 4 in Figure 5 (b). For the first one, we take the intersection CR2 of C1 and C2 which is close to the center of C3, and then we draw a straight line through CR2 and the center of C3. Next we select the intersection CR3, which is the intersection of the straight line with C3 and between CR2 and the center of C3. Finally, we take the mid-point of CR2 and CR3 as the mobile node's reference coordinates. For the second case of case 4, the two intersections of C6 with the straight line are not located between CR5 and the center of C6, so we deal with it simply and take CR5 as the mobile node's reference coordinates. The case is ignored as the three circles are without any intersections.

In Figure 5 (a) there are 10 groups of combination of any three circles, so $N=10$. After running the trilateral centroid algorithm 10 times we get 10 reference coordinates of mobile node, they are CR1, CR1, CR1, CR1, S1, S2, S3, S4, S5 and S6. We calculate the weighted average of the 10 coordinates and use the weighed average to filter out the 10 coordinates which are far away from it. For example, we filter out the reference coordinates whose distance is more than 20$cm$ with the weighted average. After the filtration there are $N_2$ ($N_2 \leq 10 \leq 10$) reference coordinates and the mobile node's coordinates are the weighted average of the $N_2$ reference coordinates.

IV. ALGORITHM DESIGN

The definitions of the symbols used in NTCWLA are shown in Table I.

TABLE I. SYMBOL DEFINITIONS

| Symbol | Description |
|---|---|
| $PTS$ | Number of beacon nodes |
| $RPN$ | $RPN$ is the largest number of the saved data packets and the packets are received from a beacon node during a localization period. Its value is affected by many factors and the main factor is the move rate of mobile node. With the increase of the rate, the value of $RPN$ is decreased. High rate causes fast location change and fast location change decreases the significance of the historical RSSI. |
| $rxDBm_{ij}$ | The RSSI of the $j$th historical packet from beacon node $i$ in a localization period. |
| $wHD_i$ | Record whether the mobile node receives more than $RPN$ packets from beacon node $i$. |
| $RR$ | A condition to filter reliable beacon node and the current RSSI of reliable beacon node must be bigger than it. |
| $MR$ | A condition to filter RSSI, when mobile node receives a packet it requires to check whether the RSSI of the packet is bigger than $MR$, if it is bigger than $MR$ then store it, else discard it. |
| $ann$ | Record the number of reliable beacon node in a localization period. |
| $crosses_{ij}$ | Record the intersection between circles, for example, $crosses_{ij}$ and $crosses_{ji}$ are the intersections of the two circles, one's center is the location of beacon node $i$ and the other's center is the location of beacon node $j$, the radiuses of them are the measured distances between mobile node. |
| $isHI_{ij}$ | Record the relationship between circles, for example, $isHI_{ij}$ ($i<j$) records the two circles' relationship, one's center is the location of beacon node $i$ and the other's center is the location of beacon node $j$. If the two circles have two intersections, the value is 1; if have one intersection, the value is 0, otherwise the value is -1. |
| $N$ | The execute times of trilateral centroid algorithm, its value is *ann * (ann-1) * (ann-2)/6*. |
| $newSite_i$ | The *i*th reference coordinates of mobile node. |
| $mr_i$ | The smallest distance between the three beacon nodes and the mobile node, and the three beacon nodes are a combination takes part in trilateral centroid localization. It is corresponding to $newSite_i$. |
| $weight_i$ | The weight of $newSite_i$. |
| $RSSI$ | Current RSSI of beacon nodes. |
| $avr\_dis_i$ | The distance between beacon node *i* and mobile node. |
| $dis\_index_i$ | Record the index of the beacon node which is identified as reliable beacon node. For example, the value of $dis\_index_0$ is *j* means the beacon node *j* is the first one that is identified as reliable beacon node. |
| $mNS$ | The measured location of mobile node. |

**Current RSSI of beacon node:**

The current RSSI of beacon node is the weighted average of multiple groups of historical RSSI, if beacon node *i* is a reliable beacon node then the current RSSI of beacon node *i* is

$$rssi_0 = \frac{rxDBm_{i0}}{2^{RPN-1}} \quad (6)$$

$$rssi_j = \frac{rxDBm_{ij}}{2^{RPN-j}} (0 < j < RPN) \quad (7)$$

$$RSSI = rssi_0 + \sum_{j=1}^{RPN-1} rssi_j \quad (8)$$

**Weight of reference coordinates of mobile node:**

After executing the trilateral centroid algorithm $N$ times there are $N_1$ ($N_1 \leq N$) reference coordinates of mobile node and the weight of the *i*th reference coordinates is

$$weight_i = \frac{\frac{1}{mr_i}}{\sum_{k=0}^{N_1-1}\frac{1}{mr_k}} \quad (9)$$

**Coordinates of mobile node:**

With $N_1$ reference coordinates of mobile node and the mobile node's coordinates are

$$mNS = \sum_{i=0}^{N_1-1} newSite_i \cdot weight_i \quad (10)$$

*A. RSSI Storing and Distance Calculation*

For localization we need to filter and store RSSIs of the packets from all beacon nodes, and then use multiple groups of historical RSSI to calculate current RSSI. Last we calculate the distances between beacon nodes and mobile node based on the calculated current RSSIs using an empirical formula. Algorithm 1 is the method of store RSSI.

**Algorithm 1: RSSI Storing**

Begin
  Step 1. Receive signal and the RSSI is *rxdbm*, signal from beacon node *i* and is the *j*th packet, judge whether *rxdbm>MR* (a threshold, used to filter RSSI), if not end, otherwise go on.
  Step 2. Judge whether *j<RPN*, if not go to Step 4, otherwise go on.
  Step 3. $rxDBm_{ij}$ =*rxdbm*, *j*++, judge whether *j==RPN*, if not go to Step 5, otherwise $wHD_i$ = *true* and go to Step 5.
  Step 4. First $rxDBm_{il}$ = $rxDBm_{i(l+1)}$ (0≤*l*<RPN-1), then $rxDBm_{i(RPN-1)}$ = *rxdbm*.
  Step 5. Output $wHD_i$ and $rxDBm_{ij}$.
End

In Algorithm 1 there is only one loop in Step 4, so the complexity of the algorithm is determined by the value of *RPN* and it is *O(n)*. Constrained by the communication rate of nodes and localization requires real time, so the value of *RPN* won't be very large and the cost of time of this algorithm is very low.

Current RSSIs are calculated at the end of the localization period and they are used to calculate the distances between mobile node and beacon nodes, as specified in Algorithm 2.

**Algorithm 2: Distance Calculation**

Begin
  Step 1. *i=0, ann=0*.
  Step 2. Judge whether *i<PTS* (number of beacon nodes), if not go to Step 3, otherwise go on.
    Step 2.1. Judge whether $wHD_i$ ==*true*, if not go to Step 2.9, otherwise go on.
    Step 2.2. *RSSI=0.0, j=0*.
    Step 2.3. Judge whether *j<RPN*, if not go to Step 2.7, otherwise go on.
    Step 2.4. Judge whether *j==0*, if not go to Step 2.6, otherwise goes on.
    Step 2.5. Based on (6), $RSSI += \frac{rxDBm_{ij}}{2^{RPN-1}}$, *j*++, go to Step 2.3.
    Step 2.6. Based on (7), $RSSI += \frac{rxDBm_{ij}}{2^{RPN-j}}$, *j*++, go to Step 2.3.
    Step 2.7. Judge whether *RSSI >RR*, if not go to Step

2.9, otherwise go on.

Step 2.8. Put *RSSI* into (3) to calculate the distance between beacon node *i* and mobile node, and then save the measured distance in $avr\_dis_{ann}$, $dis\_index_{ann}=i$, $ann++$.

Step 2.9. $i++$, go to Step 2.

Step 3. Output $avr\_dis_l$, $dis\_index_l$, $ann$, where $0 \leq l < ann$.
End

In Algorithm 2 there is a double loop and the times of loop are *PTS*, *RPN*, respectively, so the complexity of this algorithm is $O(n^2)$. Constrained by the communication rate of nodes and localization requires real time, so the value of *RPN* won't be very large. *PTN* is the number of all beacon nodes and in actual the number is limited, so the cost of time of this algorithm is also very low.

In the course of the implementation of algorithm 2 we make a record based on the *ann* for adjusting localization period dynamically, please see Figure 4 for the detailed information.

B. *N-times Trilateral Centroid Weighted Localization Algorithm*

Before NTCWLA we need to construct some circles. Circles' centers are the locations of reliable beacon nodes and circles' radiuses are the measured distances. The number of circles is *ann*. After that we require to judge the relationship of any two circles and at the same time save the intersections of them. Next combine any three reliable beacon nodes to calculate the reference coordinates of mobile node with trilateral centroid algorithm. Finally, we deal with the reference coordinates to locate the mobile node. Detailed information is shown in Algorithm 3 as follows:

**Algorithm 3:** NTCWLA

Begin
Step 1. $i=0$.
Step 2. Judge whether $i<ann-1$, if not go to Step 3, otherwise go on.
Step 2.1. $j=i+1$.
Step 2.2. Judge whether $j<ann$, if not go to Step 2.5, otherwise go on.
Step 2.3. Check the relationship and calculate intersections of two circles whose center is the positions of beacon nodes $dis\_index_i$ and $dis\_index_j$, and then record them in $isHI_{ij}$, $crosses_{ij}$ and $crosses_{ji}$, respectively.
Step 2.4. $j++$, go to Step 2.2.
Step 2.5. $i++$, go to Step 2.
Step 3. $i=0$, $l=0$.
Step 4. Judge whether $i<ann-2$, if not go to Step 5, otherwise go on.
Step 4.1. $j=i+1$.
Step 4.2. Judge whether $j<ann-1$, if not go to Step 4.7, otherwise go on. Step 4.3: $k=j+1$.

Step 4.4. Judge whether $k<ann$, if not go to Step 4.6, otherwise go on.
Step 4.5. Check the value of $isHI_{ij}$, $isHI_{jk}$ and $isHI_{ik}$, and then judge the relationship of the three circles whose centers are the positions of beacon nodes $dis\_index_i$, $dis\_index_j$ and $dis\_index_k$, respectively. Use trilateral centroid algorithm to calculate the mobile node's reference coordinates and save it in $newSite_l$, select the smallest one from $avr\_dis_i$, $avr\_dis_j$, $avr\_dis_k$ and then save it in $mr_l$. $k++$, $l++$, go to Step 4.4.
Step 4.6. $j++$, go to Step 4.2.
Step 4.7. $i++$, go to Step 4.
Step 5. Based on (9), the weight of reference coordinates $newSite_i$ is calculated.
Step 5.1. Based on (10) calculate the mobile node's coordinates first time (*mNS* is calculated location of mobile node).
Step 5.2. Filter out the reference coordinates of the mobile node which are far away from the value which calculated in Step 5.1, and then based on (10) to locate the mobile node second time with the filtered reference coordinates.
Step 6. Output *mNS*.
End

In Algorithm 3 there is a triple loop and the times of loop are *ann*, *ann*-1, *ann*-2, respectively, so the complexity of this algorithm is $O(n^3)$. The reliable beacon node is limited so that the cost of time of this algorithm is low.

V. EXPERIMENTS AND RESULTS

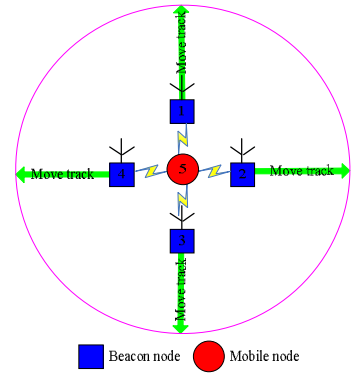

Figure 6. Topology for getting empirical formula

We experiment in indoor with the chip STM32W108 provided by company ST. This kind of chip with 32-bit ARM Cortex core and supports IEEE 802.15.4 protocol. The experimental area is a place of $1m \times 1m$. There is a mobile node in all experiments and other nodes are as beacon node. The mobile node first creates a wireless network and then all beacon nodes join it in turn. After that mobile node receives signals and saves the corresponding

RSSIs. All beacon nodes' send-power is the same and ideally if the distances between mobile node and beacon nodes are the same the RSSIs of the beacon nodes will be the same. However, in actual experiments the RSSIs of the beacon nodes are not the same due to the differences of hardware and many other external factors.

Before the localization experiments we need to ensure the parameters of Formula (3) to get the empirical formula, Figure 6 is the network topology to get empirical formula.

There are five nodes in Figure 6. The node 5 is mobile node and other nodes are beacon nodes. The mobile node receives signals from all beacon nodes and transports them to PC through serial port. PC filters out signals whose strength is less than -70*dbm*, and then saves the RSSIs and the corresponding distances of beacon nodes. The 4 beacon nodes move along 4 radiuses of a circle whose radius is 1.5*m* and the angle of any two adjacent radiuses is 90°. The moving distance interval of beacon nodes is 10*cm* and time interval is 1*min*. All beacon nodes start communicating with mobile node from the sites, where are 10*cm* away from the mobile node. Then we move all beacon nodes 10*cm* away every minute along the radiuses until out of the test area. After data collection, we deal with the saved data to get the parameters of Formula (3). During this process we adjust *smooth* to get different values of $p_1$ and $p_2$. Last we take the average values of the measured RSSI into the empirical formula to get the measured distances and the results are shown in Table II. Table III is the measured error corresponding to Table II.

TABLE II. ACTUAL DISTANCES AND MEASURED DISTANCES

| Actual distances (*cm*) | Measured distances (*cm*) | | | | |
|---|---|---|---|---|---|
| | *smooth*=1 p1=-11.17 p2=6.037 | *smooth*=3 p1=-11.32 p2=6.769 | *smooth*=7 p1=-11.36 p2=7.148 | *smooth*=9 p1=-11.35 p2=7.178 | *smooth*=11 p1=-11.33 p2=7.176 |
| 10 | 12.87 | 13.27 | 13.60 | 13.66 | 13.73 |
| 30 | 28.57 | 29.16 | 29.79 | 29.96 | 30.14 |
| 50 | 41.73 | 42.37 | 43.24 | 43.50 | 43.78 |
| 70 | 57.51 | 58.14 | 59.26 | 59.63 | 60.05 |
| 90 | 87.43 | 87.91 | 89.47 | 90.06 | 90.77 |
| 110 | 110.38 | 110.63 | 112.51 | 113.28 | 114.21 |
| 130 | 130.05 | 130.07 | 132.20 | 133.12 | 134.25 |
| 150 | 185.98 | 185.13 | 187.93 | 189.29 | 191.02 |

TABLE III. ACTUAL DISTANCES AND MEASURED ERRORS

| Actual distances (*cm*) | Measured error (*cm*) | | | | |
|---|---|---|---|---|---|
| | *smooth*=1 p1=-11.17 p2=6.037 | *smooth*=3 p1=-11.32 p2=6.769 | *smooth*=7 p1=-11.36 p2=7.148 | *smooth*=9 p1=-11.35 p2=7.178 | *smooth*=11 p1=-11.33 p2=7.176 |
| 10 | 2.87 | 3.27 | 3.60 | 3.66 | 3.73 |
| 30 | -1.42 | -0.83 | -0.20 | -0.03 | 0.14 |
| 50 | -8.26 | -7.62 | -6.75 | -6.49 | -6.21 |
| 70 | -12.48 | -11.85 | -10.73 | -10.36 | -9.94 |
| 90 | -2.56 | -2.0859 | -0.52 | 0.06 | 0.77 |
| 110 | 0.38 | 0.6399 | 2.51 | 3.28 | 4.21 |
| 130 | 0.05 | 0.07 | 2.20 | 3.12 | 4.25 |
| 150 | 35.98 | 35.13 | 37.93 | 39.29 | 41.02 |

It can be seen from Table III that when the actual distance between beacon node and mobile node is 70*cm* the negative deviation is maximal; when the actual distance is 150*cm* the positive deviation is maximal. The test area is 1*m*×1*m* so the largest distance between any nodes is about 140*cm* and in most case the distance is within 100*cm*. If we want to improve the localization accuracy, we require to increasing *smooth* to decrease the negative deviation. Table II and Table III show that when *smooth*=7, the value of $p_1$ is minimum and when *smooth*=9, the value of $p_2$ is maximum. Based on Formula (4) and (5), we can get the best empirical formula:

$$d = e^{(\frac{P(d)+7.163}{-11.355})} \qquad (11)$$

*A. Experiment 1*

There are 9 nodes in this experiment. One of them is mobile node and others are beacon nodes, and the locations of beacon nodes are fixed. The distribution of beacon nodes is shown in Figure 7. In this experiment the trace of the mobile node is the diagonal, and the mobile node moves from point (0.1, 0.1) to point (0.9, 0.9). Constrained by experimental conditions, the movement of mobile node is dragged by a string artificially, so the speed can't be controlled very well. And movement increases the instability of signals. In order to improve the localization accuracy, we only can increase the localization period and decrease the move rate. The localization period is 1*s* and the value of *RPN* is 5 (Note: The environment of this experiment is different from the experiment which corresponding to the Figure 1 and Figure 2, *RPN* =5 is reliable here).

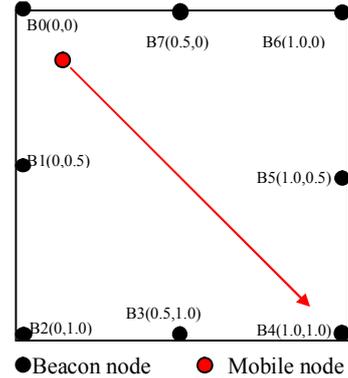

Figure 7. Network topology and linear motion trace of mobile node

The accuracy of NTCWLA is affected by the value of *N* and *N* is determined by *n*. Constrained by the beacon nodes' number and the communication rate, *n* is 3, 4, 5 and 6. When *n*=3 the NTCWLA changes into trilateral centroid algorithm. *n*=3 is used to compare with *n*=4, 5, 6. Figure 8 shows the localization results.

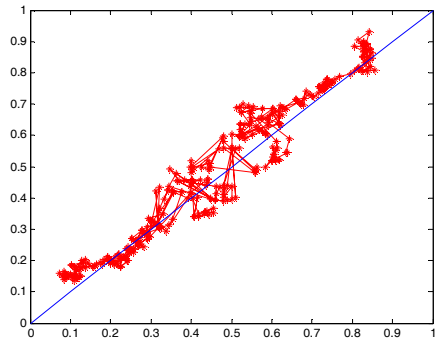

(a) n=3

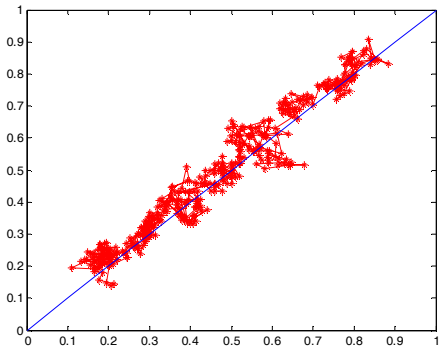

(b) n=4

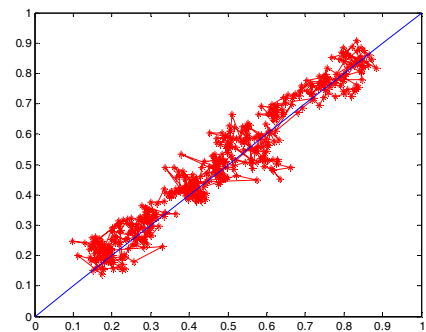

(c) n=5

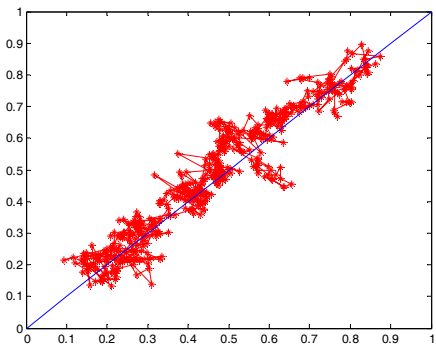

(d) n=6

Figure 8. Mobile node's localization results of linear motion

The localization results show that when $n$=5 the localization accuracy is the best and the result of $n$=6 is not as well as the result of $n$=5. $n$=6 means there are some nodes' reliability decreasing and this increases the localization error. Figure 8 (b) means less reliable beacon nodes cause poor localization accuracy. The four figures in Figure 8 show that when mobile node is in the center of the test area the localization accuracy is not very good, because all beacon nodes are far away from the mobile node and the measured RSSIs are not very accurate. The measured data in Table II and Table III can explain this.

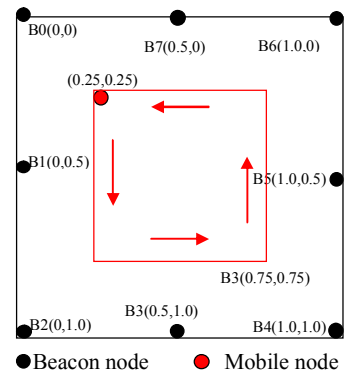

Figure 9. Network topology and curvilinear motion trace of mobile node

### B. Experiment 2

The trace in Experiment 1 is a straight line, as a contrast the trace is a curve in Experiment 2 as shown in Figure 9. The red square is the trace of mobile node and the arrows represent the directions of movement. All conditions of this experiment are the same with Experiment 1 and $n$ is 3, 4, 5 and 6. The function of $n$=3 is the same with Experiment 1. Figure 9 describes the experiment topology and Figure 10 shows the localization results.s

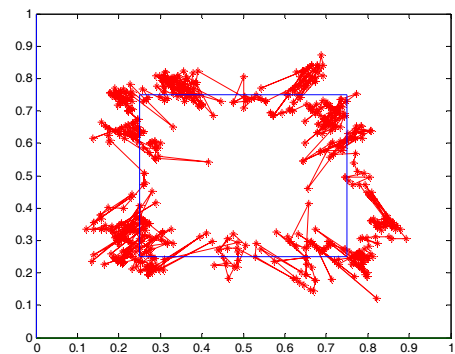

(a) n=3

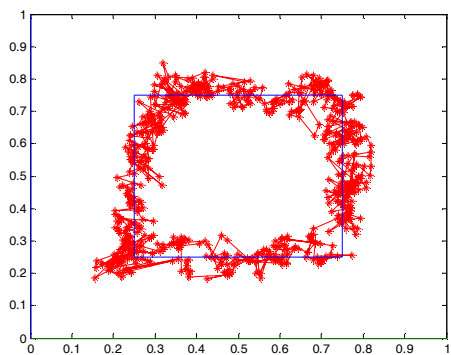

(b) n=4

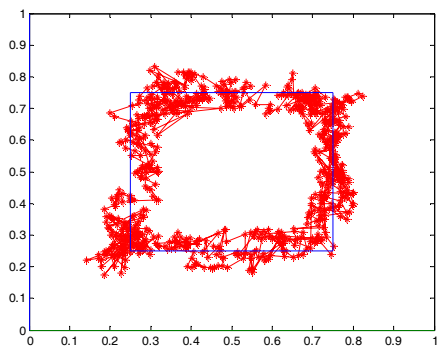

(c) n=5

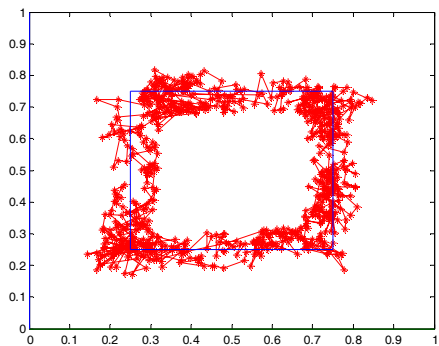

(d) n=6

Figure 10. Mobile node's localization results of curvilinear motion

The results of Experiment 2 show that when $n=4$ the localization accuracy is the best. With the increase of $n$ some reliable beacon nodes' reliability is decreasing so that the localization accuracy is decreased. The difference of Experiment 1 with Experiment 2 is that in Experiment 2 whatever the mobile node is there are always some beacon nodes which are close to it and others are far away from it. So with the increase of $n$ there will be some beacon nodes' reliability is decreasing and this will increase the localization error.

From the localization results of Experiment 1 and Experiment 2, the conclusion NTCWLA is better than that of trilateral centroid algorithm can be got. Figure 8 (a) and Figure 10 (a) show that the localization accuracy of $n=3$ is not as well as $n=4$, 5 and 6. The points are not even distributed in Figure 8 and Figure 10. One of the reasons is the mobile node's movement is controlled by hand and we can't keep it to do uniform motion. When the mobile node is far away from some beacon nodes, the localization error is great and this will lead to fewer points in some places.

## VI. CONCLUSIONS

The periodical localization of mobile node with NTCWLA is based on RSSI. The steps of the new algorithm are: First we select a number of reliable beacon nodes from large number of beacon nodes and calculate the distances between mobile node and reliable beacon nodes using an empirical formula. Then we select any three nodes from these reliable beacon nodes to calculate reference coordinates of mobile node using trilateral centroid algorithm. During this process we give each reference coordinates a weight, which is based on the distances between mobile node and the three selected reliable beacon nodes. Finally, we use the weighted average of all reference coordinates to filter the reference coordinates and take the weighted average of the filtered reference coordinates as the mobile node's location. Experiment results show that this new localization algorithm is better than trilateral centroid algorithm.

It is worth noting that $N$ affects the localization accuracy and it is determined by the number of selected reliable beacon nodes. Different traces and different topology of beacon nodes corresponding to different optimal number of selected reliable beacon nodes. This algorithm will be used in our future work on vehicle navigation system, which uses wireless sensor network to control the vehicle's movement, but the speed of the vehicle is constrained by the periodicity of localization.

ACKNOWLEDGMENT

This work was supported in part by Natural Science Foundation of China under Grant No. 60773213 and Grant No. 60903153, Program for New Century Excellent Talents in University (NCET-09-0251), the Fundamental Research Funds for the Central Universities, and the SRF for ROCS, SEM.